\documentclass[a4paper]{jpconf}
\usepackage{iopams}
\usepackage{graphicx}
\usepackage{amsmath}
\usepackage{physics}
\usepackage{microtype}
\usepackage{hyperref} \hypersetup{colorlinks=true,linktoc=all,linkcolor=blue,breaklinks=true,citecolor=blue,urlcolor=blue}
\usepackage[square,sort,numbers,compress]{natbib}
\usepackage{changes}

\begin{document}
\title{Equilibrium finite-frequency noise of an interacting mesoscopic capacitor studied in time-dependent density functional theory}

\author{Niklas Dittmann$^{1,2,3}$, Janine Splettstoesser$^{1}$ and Nicole Helbig$^{3}$}

\address{$^1$ Department of Microtechnology and Nanoscience (MC2), Chalmers University of Technology, SE-41258 G{\"o}teborg, Sweden}
\address{$^2$ Institute for Theory of Statistical Physics, RWTH Aachen, D-52056 Aachen, Germany}
\address{$^3$ Peter-Gr\"unberg Institut and Institute for Advanced Simulation, Forschungszentrum J\"ulich, D-52425 J\"ulich, Germany}

\ead{dittmann@physik.rwth-aachen.de}

\begin{abstract}
We calculate the frequency-dependent equilibrium noise of a mesoscopic capacitor in time-dependent density functional theory (TDDFT). 
The capacitor is modeled as a single-level quantum dot with on-site Coulomb interaction and tunnel coupling to a nearby reservoir. 
The noise spectra are derived from linear-response conductances via the fluctuation-dissipation theorem. 
Thereby, we analyze the performance of a recently derived exchange-correlation potential with time-nonlocal density dependence in the finite-frequency linear-response regime. % over a broad frequency range.
We compare our TDDFT noise spectra with real-time perturbation theory and find excellent agreement for noise frequencies below the reservoir temperature.
\end{abstract}

\section{Introduction}
  
The presence of thermal and quantum fluctuations in mesoscopic conductors results in noise, even if the mean current through a sample vanishes \cite{Blanter00,Clerk10}.
The noise spectrum gives insight into temporal correlations between fluctuation events. 
Measured in equilibrium\footnote{
For systems which are out of equilibrium, the noise includes information which is not accessible by conductance measurements alone, see e.\,g.~the experiments reported in Refs.~\cite{Parmentier12,Ubbelohde12}.}, 
it reveals dissipative properties of a system, i.\,e.~the real part of the finite-frequency linear-response conductance, by means of the fluctuation-dissipation theorem \cite{Callen51}.

In this paper, we present the calculation of equilibrium finite-frequency noise spectra in time-dependent density functional theory (TDDFT) \cite{MarquesMaitraNogueiraGrossRubio12,Ullrich11,Maitra16}.
In TDDFT, the dynamics of an interacting quantum system is simulated by a non-interacting Kohn-Sham (KS) system \cite{Runge84}, which leads to cost-efficient numerics even for complex systems.
The KS potential, which consists of the external potential of the interacting system, modified by the Hartree (H) and the exchange-correlation (XC) contributions, is chosen such that the KS and the interacting
systems share the same time-dependent electronic density. The H potential accounts for electrostatics, while all further many-body effects are included in the universal XC potential.
The development and characterization of approximations for the latter is among the central challenges for TDDFT, and also a focus of this work.

\begin{figure}[t]
\begin{center}
\includegraphics[width=0.98\columnwidth]{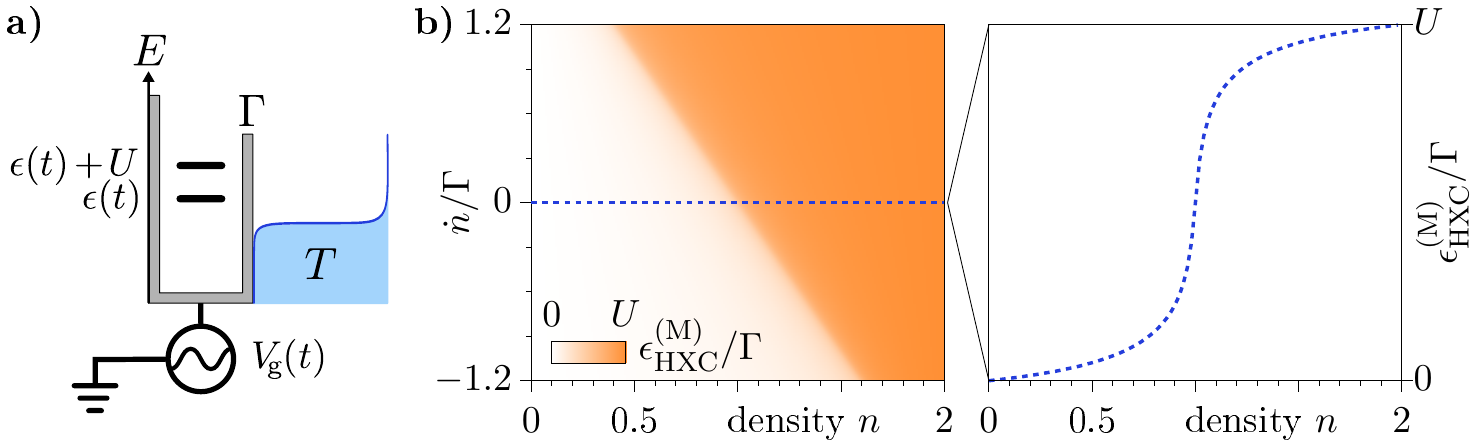}
\end{center}
\vspace{-5mm}
    \caption{(a) Energy diagram of a single-level quantum dot with tunnel coupling, $\Gamma$, to a reservoir; (b)
    non-adiabatic HXC potential, $\epsilon_{\mathrm{HXC}}^{(\mathrm{M})}$, see Ref.~\cite{Dittmann17}, for $U=16\Gamma$, $T=2\Gamma$; 
    dotted line marks the line cut shown on the r.\,h.\,s.}
    \label{fig_HXC}
\end{figure}

As an instructive system with experimental relevance \cite{Feve07}, we analyze an interacting mesoscopic capacitor.
For this system, the finite-frequency noise has been studied theoretically, e.\,g.~\cite{Buettiker92,Rothstein09,Gabdank11}, and in experiment \cite{Parmentier12}.
We model the interacting mesoscopic capacitor by a single-level quantum dot with on-site Coulomb interaction and tunnel coupling to an adjacent reservoir, see Fig.~\ref{fig_HXC} (a). 
Based on this model, we recently developed a non-adiabatic, i.\,e.~time nonlocal, XC potential for TDDFT simulations of single-electron tunneling devices \cite{Dittmann17}.
The XC potential was derived by expressing the dynamics of both the system, and its KS counterpart, in terms of time-dependent Markovian\footnote{
The Markov approximation neglects memory in the reservoir. The latter becomes relevant for dynamics on time scales smaller than the reservoir memory time, which
is given by the inverse of the reservoir temperature $T$.} Master equations.
It thereby includes information on the dynamics of electron tunneling.
In Ref.~\cite{Dittmann17}, we applied the non-adiabatic XC potential to describe the operation mode of a mesoscopic capacitor as a single-electron source, i.\,e.~its non-linear response to a time-periodic gate voltage.
A key finding was that a related \emph{adiabatic} XC potential leads to electron dynamics on time scales which correspond to non-interacting electrons,
while our \emph{non-adiabatic} XC potential describes electron dynamics on the respective time scales for interacting electrons.
Remarkably, our non-adiabatic XC potential thereby only depends on the time-local density of the quantum dot and its first time derivative.

In the present paper, we add an analysis of finite-frequency linear-response physics to our previous discussion.
For this purpose, we derive finite-frequency conductances of the system at hand with TDDFT. 
We then consider the real parts of the obtained conductances, which we express as equilibrium noise by means of the fluctuation-dissipation theorem.
To test the performance of our non-adiabatic XC potential over a broad frequency range, we compare our results with noise spectra derived with perturbation theory.

The paper is structured as follows: We define our model for the interacting mesoscopic capacitor in Sec.~\ref{sec_model} and its KS counterpart in Sec.~\ref{sec_modelKS}, 
where we also review the non-adiabatic XC potential from Ref.~\cite{Dittmann17}. The calculation of equilibrium noise in TDDFT is described in Sec.~\ref{sec_calc}.
In Sec.~\ref{sec_discussion} we present TDDFT noise spectra, before we conclude our work in Sec.~\ref{sec_conclusion}.

%%%%%%%%%%%%%%%%%%%%%%%%%%%%%%%%%%%%%%%%%%%%%%%%%%%%%%%%%%%%%%%%%%%%%%%%%%%%%%
% figure: setup+HXC
%%%%%%%%%%%%%%%%%%%%%%%%%%%%%%%%%%%%%%%%%%%%%%%%%%%%%%%%%%%%%%%%%%%%%%%%%%%%%%
%%%%%%%%%%%%%%%%%%%%%%%%%%%%%%%%%%%%%%%%%%%%%%%%%%%%%%%%%%%%%%%%%%%%%%%%%%%%%%

%%%%%%%%%%%%%%%%%%%%%%%%%%%%%%%%%%%%%%%%%%%%%%%%%%%%%%%%%%%%%%%%%%%%%%%%%%%%%%
\section{The interacting mesoscopic capacitor}
\label{sec_model}

We model the mesoscopic capacitor by a quantum dot with a single energy level, which is tunnel coupled to a nearby reservoir, see Fig.~\ref{fig_HXC} (a).
Therefore, the dot can be either empty, singly or doubly occupied.
The energy level, $\epsilon(t)=-\alpha V_\mathrm{g}(t)$, is assumed to depend linearly on the gate voltage, $V_\mathrm{g}(t)$, with $\alpha > 0$. 
All energies are defined with respect to the Fermi energy and $e,\hbar$ and $k_\mathrm{B}$ are set to one.
We consider strong Coulomb repulsion between electrons occupying the quantum dot, by taking into account a charging energy $U$ in the case of double occupation.
In addition, we assume the electrons in the reservoir to be non-interacting, set the electrochemical potential to zero, and denote the reservoir temperature by $T$.
The resulting Hamiltonian reads
\begin{align}
\label{eq_hamiltonian}
 H &= \sum_\sigma \epsilon(t) d_\sigma^\dagger d_\sigma + U d_\uparrow^\dagger d_\uparrow d_\downarrow^\dagger d_\downarrow + 
     \sum_{k,\sigma} \epsilon_k c_{k\sigma}^\dagger c_{k\sigma} + \left(\gamma \sum_{k,\sigma} c_{k\sigma} d_{\sigma}^\dagger +  \mathrm{h.\,c.}\right),
\end{align}
where $d^{(\dagger)}_\sigma$ denote the annihilation (creation) operators for quantum-dot states with spin index $\sigma ={} \uparrow,\downarrow$ and $c^{(\dagger)}_{k\sigma}$ denote
the annihilation (creation) operators for reservoir states with spin $\sigma$ and momentum $k$.
We assume spin-degeneracy and consider a single energy band, $\epsilon_k$, %, assume spin degeneracy
and an energy-independent tunnel coupling, $\gamma$. Besides that, we define $\Gamma = 2\pi |\gamma|^2 \nu_0$
as the tunnel-coupling strength, where $\nu_0$ denotes the density-of-states (DOS) at the Fermi energy.
The density on the quantum dot is given by $n(t) = \sum_\sigma \expval{d^\dagger_\sigma d_\sigma}(t)$, and the current into the quantum dot reads $I(t)=-\dot{n}(t)$ due to charge conservation.

%%%%%%%%%%%%%%%%%%%%%%%%%%%%%%%%%%%%%%%%%%%%%%%%%%%%%%%%%%%%%%%%%%%%%%%%%%%%%%
\section{Representation by a non-interacting system}
\label{sec_modelKS}

We now describe the interacting mesoscopic capacitor defined in Eq.~\eqref{eq_hamiltonian} by a non-interacting KS system, assuming non-interacting $v$-representability \cite{MarquesMaitraNogueiraGrossRubio12,Ullrich11,Maitra16}. 
The H and XC potentials are taken into account via a combined shift of the quantum-dot's energy level by the amount $\epsilon_\mathrm{HXC}[n](t)$, which we call the HXC potential of our model system.
By furthermore setting the on-site interaction in Eq.~\eqref{eq_hamiltonian} to zero, we obtain the KS Hamiltonian,
\begin{align}
\label{eq_hamiltonianKS}
 H_\mathrm{KS} &= \sum_\sigma \epsilon_\mathrm{KS}[n](t) d_\sigma^\dagger d_\sigma + 
     \sum_{k,\sigma} \epsilon_k c_{k\sigma}^\dagger c_{k\sigma} + \left(\gamma \sum_{k,\sigma} c_{k\sigma} d_{\sigma}^\dagger +  \mathrm{h.\,c.}\right),
\end{align}
with the KS energy level, $\epsilon_\mathrm{KS}[n](t) = \epsilon(t) + \epsilon_\mathrm{HXC}[n](t)$, and where a general functional dependence on the density is denoted by square brackets, $[n]$.
For the HXC potential in Eq.~\eqref{eq_hamiltonianKS}, we analyze the non-adiabatic approximation derived in Ref.~\cite{Dittmann17}, $\epsilon_\mathrm{HXC}^{\mathrm{(M)}}(n(t),\dot{n}(t))(t)$,
and its adiabatic counterpart, $\epsilon_\mathrm{HXC}^{\mathrm{(A)}}(n(t))(t)$, see also Ref.~\cite{Stefanucci11}. 
The two approximations read
\begin{align}
\begin{aligned}
 \label{eq_hxcMA}
 \epsilon_\mathrm{HXC}^{\mathrm{(M)}}(n(t),\dot{n}(t))(t) &= \frac{1}{\beta} \log \left(C(n(t),\dot{n}(t))\right)
\end{aligned}
\end{align}
and $\epsilon_\mathrm{HXC}^{\mathrm{(A)}}(n(t))(t) = \epsilon_\mathrm{HXC}^{\mathrm{(M)}}(n(t),0)(t)$, with the inverse temperature, $\beta = 1/T$ and
\begin{align}
 \frac{1}{C(n,\dot{n})} &= \frac{\dot{n} + e^{U\beta}\left(\dot{n}+2\Gamma (n-1)\right)}{2 e^{U\beta}(\dot{n}+\Gamma(n-2))}\Bigg(1 - 
 \Bigg(1-\frac{4e^{U\beta}\left((\dot{n}+\Gamma n)^2-2\Gamma (\dot{n}+\Gamma n)\right)}{\left(\dot{n} +e^{U\beta}\left(\dot{n}+2\Gamma (n-1)\right)\right)^2}\Bigg)^{\frac{1}{2}}\Bigg).
\end{align}
The adiabatic HXC potential is presented in the line cut in Fig.\ \ref{fig_HXC} (b), where we observe a smeared-out step in the potential at single occupation of the quantum dot \cite{Stefanucci11,Evers11}.
This step is known in TDDFT and DFT as the derivative discontinuity \cite{Perdew82}, 
which is, e.\,g., relevant for a proper description of Coulomb-blockade physics in a non-interacting Kohn-Sham system \cite{Evers11,Kurth10}.
The density plot in Fig.~\ref{fig_HXC} (b), showing the non-adiabatic HXC potential, reveals that a non-zero time derivative of the density shifts this step to a different position.
The physical interpretation of this shift is explained in detail in Ref.~\cite{Dittmann17}: its magnitude is linked to the difference of relaxation times of the interacting and the non-interacting system.
The accomplishment of this dynamical step is an improved description of the relaxation dynamics of the \emph{interacting} mesoscopic capacitor by the \emph{non-interacting} KS system. 
This is not achieved by considering the adiabatic HXC potential only, where relaxation occurs on a time scale which is characteristic for a non-interacting mesoscopic capacitor \cite{Dittmann17}.

%%%%%%%%%%%%%%%%%%%%%%%%%%%%%%%%%%%%%%%%%%%%%%%%%%%%%%%%%%%%%%%%%%%%%%%%%%%%%%
\section{Equilibrium finite-frequency noise in TDDFT}
\label{sec_calc}

Within TDDFT, we calculate the equilibrium finite-frequency noise, $S(\omega)$, of the interacting mesoscopic capacitor,
which we define as the Fourier transform of the symmetric current-current fluctuations, $S(t-t') = \expval{I(t)I(t')+I(t')I(t)}$.
In this work, we extract the noise spectrum from the linear-response conductance $G(\omega) = \left.\frac{\partial I(\omega)}{\partial \epsilon(\omega)}\right|_\mathrm{eq}$,
where the subscript denotes evaluation at the equilibrium density, $n_\mathrm{eq}$.
This is possible due to the fluctuation-dissipation theorem,
\begin{align}
 \label{eq_fdt}
 S(\omega) &= 2 \omega \coth\left(\frac{\beta\omega}{2}\right)\Re G(\omega),
\end{align}
which explicitly links the dissipative part of the conductance to the equilibrium noise \cite{Blanter00}.
To obtain $G(\omega)$ in TDDFT, we first calculate the conductance in the KS system, which is defined as
$G_\mathrm{KS}(\omega) = \left.\frac{\partial I(\omega)}{\partial \epsilon_\mathrm{KS}[n](\omega)}\right|_\mathrm{eq}$.
The latter depends on the position of the KS energy level, $\epsilon_\mathrm{KS}[n_\mathrm{eq}]$, and is derived exactly, because the KS system is non-interacting, see e.\,g.~the result in Ref.~\cite{Jauho94}.
The density $n_\mathrm{eq}$ is calculated self-consistently from the exact KS expression for the quantum-dot's equilibrium density,
$n_\mathrm{eq} = \frac{\Gamma}{\pi} \int dq \frac{ d_q}{\left(\epsilon_\mathrm{KS}[n_\mathrm{eq}]-q\right)^2+\Gamma^2/4}$,
with the Fermi function, $d_q = 1/(1+e^{\beta q})$.
In TDDFT linear-response theory, the KS conductance is related to the conductance of the interacting system by a Dyson equation. This leads to
\begin{align}
 \label{eq_Gdyson}
  G(\omega) &= \frac{\omega G_\mathrm{KS}(\omega)}{\omega+i f_\mathrm{HXC}(n_\mathrm{eq},\omega)G_\mathrm{KS}(\omega)},
\end{align}
with the frequency-dependent HXC kernel, $f_\mathrm{HXC}(n_\mathrm{eq},\omega)$, defined as the Fourier transform of $f_\mathrm{HXC}(n_\mathrm{eq},t-t') = \left.\frac{\delta \epsilon_\mathrm{HXC}[n](t)}{\delta n(t')}\right|_\mathrm{eq}$,
see e.\,g.~\cite{MarquesMaitraNogueiraGrossRubio12,Ullrich11,Maitra16}.
%%%%%%%%%%%%%%%%%%%%%%%%%%%%%%%%%%%%%%%%%%%%%%%%%%%%%%%%%%%%%%%%%%%%%%%%%%%%%%
% figure: fHXC
%%%%%%%%%%%%%%%%%%%%%%%%%%%%%%%%%%%%%%%%%%%%%%%%%%%%%%%%%%%%%%%%%%%%%%%%%%%%%%
\begin{figure}[b]
\begin{center}
\includegraphics[width=0.75\columnwidth]{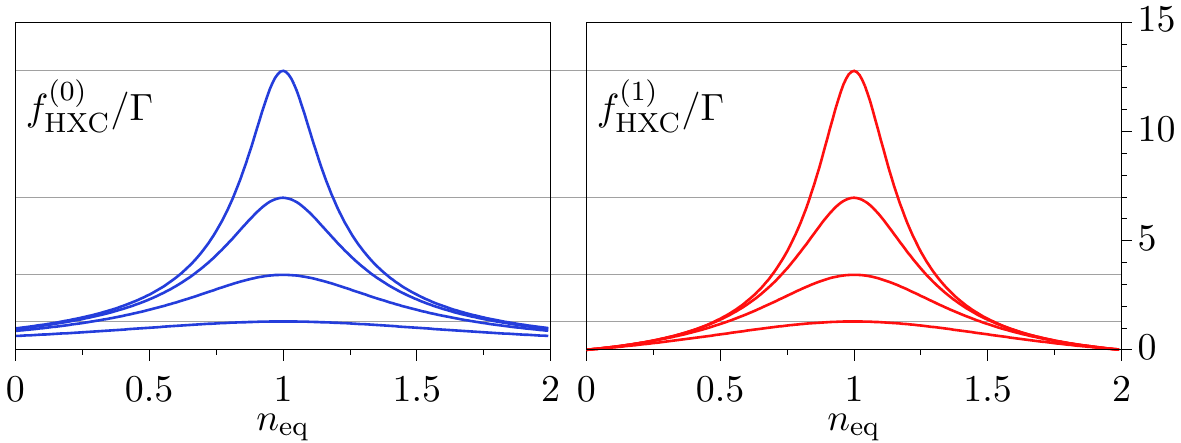}
\end{center}\vspace{-5mm}
	\caption{
	Coefficients of the HXC kernels in Eq.~\eqref{eq_fhxc} for on-site interaction $U = 2\Gamma,4\Gamma,6\Gamma,8\Gamma$  (lower to upper lines) and
	$T=2\Gamma$. Thanks to the characteristic time scale $\tau_\mathrm{ch}$ (defined in the main text), the peak heights are identical for both coefficients, while differences appear for $n_\mathrm{eq} \rightarrow 0$ and $n_\mathrm{eq} \rightarrow 2$.
	}
	\label{fig_fHXC}  
\end{figure}
%%%%%%%%%%%%%%%%%%%%%%%%%%%%%%%%%%%%%%%%%%%%%%%%%%%%%%%%%%%%%%%%%%%%%%%%%%%%%%
The HXC kernels related to our specific approximations in Eqs.~\eqref{eq_hxcMA} read
\begin{align}
\begin{aligned}
 \label{eq_fhxc}
 f_\mathrm{HXC}^{(\mathrm{M})}(n_\mathrm{eq},\omega) &= f_\mathrm{HXC}^{(0)}(n_\mathrm{eq}) - i \omega \tau_\mathrm{ch} f_\mathrm{HXC}^{(1)}(n_\mathrm{eq})
\end{aligned}
\end{align}
and $f_\mathrm{HXC}^{(\mathrm{A})}(n_\mathrm{eq}) = f_\mathrm{HXC}^{(\mathrm{M})}(n_\mathrm{eq},0)$,	
with $f_\mathrm{HXC}^{(0)}(n_\mathrm{eq}) = \frac{\partial \epsilon^{(\mathrm{M})}_\mathrm{HXC}(n,\dot{n})}{\partial n}\Big|_\mathrm{eq}$ and 
$f_\mathrm{HXC}^{(1)}(n_\mathrm{eq}) = \tau_\mathrm{ch}^{-1}\frac{\partial \epsilon^{(\mathrm{M})}_\mathrm{HXC}(n,\dot{n})}{\partial \dot{n}}\Big|_\mathrm{eq}$.
Here, we introduce a characteristic time scale, $\tau_\mathrm{ch}$, which for our approximations is convenient to define as $\tau_\mathrm{ch} = \frac{1}{\Gamma} - \frac{1}{\Gamma_\mathrm{c}}$, with the 
charge relaxation rate at the electron-hole symmetric point, $\Gamma_\mathrm{c} = \Gamma\cdot \left(1+d_{-U/2}-d_{U/2}\right)$.
The scale $\tau_\mathrm{ch}$, which also quantifies the dynamical step visible in Fig.~\ref{fig_HXC} (b) \cite{Dittmann17}, is thus given 
by the difference between the charge relaxation times of the non-interacting and the interacting system.
Fig.~\ref{fig_fHXC} shows that both real coefficients, which appear on the r.\,h.\,s.~in Eq.~\eqref{eq_fhxc}, develop a peak at single occupation for increasing on-site interaction $U$.
Therefore, the frequency dependence of $f_\mathrm{HXC}^{(\mathrm{M})}$ adds a linear and imaginary contribution to the peak of the frequency-independent HXC kernel, $f_\mathrm{HXC}^{(\mathrm{A})}$.

%%%%%%%%%%%%%%%%%%%%%%%%%%%%%%%%%%%%%%%%%%%%%%%%%%%%%%%%%%%%%%%%%%%%%%%%%%%%%%
\section{TDDFT noise spectra of the interacting mesoscopic capacitor}
\label{sec_discussion}

%%%%%%%%%%%%%%%%%%%%%%%%%%%%%%%%%%%%%%%%%%%%%%%%%%%%%%%%%%%%%%%%%%%%%%%%%%%%%%
% figure: noise
%%%%%%%%%%%%%%%%%%%%%%%%%%%%%%%%%%%%%%%%%%%%%%%%%%%%%%%%%%%%%%%%%%%%%%%%%%%%%%
\begin{figure}[b]
\begin{center}
\includegraphics[width=0.98\columnwidth]{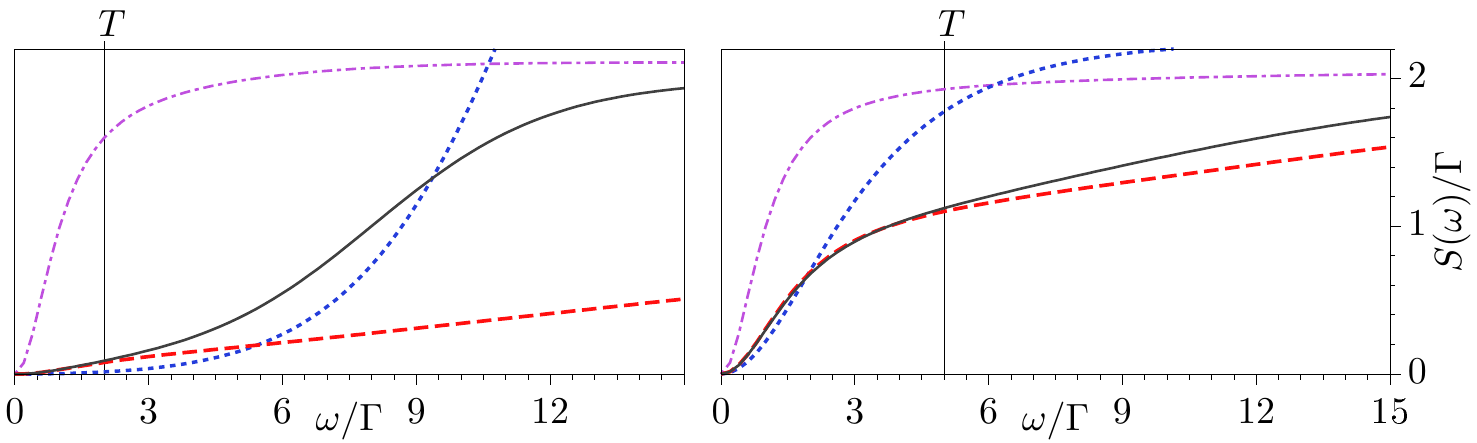}
\end{center}\vspace{-5mm}
	\caption{Equilibrium noise spectra for two temperatures, $T=2\Gamma$ (left) and $T=5\Gamma$ (right);
	Shown are the uncorrected noise spectra of the KS systems (purple dashed-dotted lines), and the corrected noise spectra (Eq.~\eqref{eq_Gdyson}), 
	using the HXC kernels $f_\mathrm{HXC}^{(\mathrm{A})}$ (blue dotted lines) and $f_\mathrm{HXC}^{(\mathrm{M})}$ (red dashed lines). 
	Further parameters are $\epsilon = -8 \Gamma, U = 16 \Gamma$.
	The black solid lines are shown for reference and present results obtained with perturbation theory \cite{Braun06,Droste15}.
	}
	\label{fig_noise}  
\end{figure}
%%%%%%%%%%%%%%%%%%%%%%%%%%%%%%%%%%%%%%%%%%%%%%%%%%%%%%%%%%%%%%%%%%%%%%%%%%%%%%

\noindent Equilibrium noise spectra of the interacting mesoscopic capacitor, using Eq.~\eqref{eq_fdt}, are plotted in Fig.~\ref{fig_noise} for two different temperatures (left and right).
As a reference, the black solid lines present the result of an analytic calculation based on perturbation theory, see e.\,g.~Refs.~\cite{Braun06,Droste15}.
This method applies an expansion in the tunnel coupling. 
For our system, the noise derived with perturbation theory features smeared-out steps at noise frequencies which equal the absolute value of $\epsilon$ or $\epsilon+U$.
Besides that, the noise tends to zero in the zero-frequency limit.
The single step for the parameters in Fig.~\ref{fig_noise} is readily visible in the left plot by the black solid line, and strongly smeared-out in the right plot due to the increased temperature.
In both plots in Fig.~\ref{fig_noise}, three TDDFT curves are shown, which respectively correspond to the noise in the KS system, obtained from $G_\mathrm{KS}(\omega)$,
and the noise calculated from the corrected conductances, see Eq.~\eqref{eq_Gdyson}, using the two HXC kernels defined in Eq.~\eqref{eq_fhxc}.
For both temperatures, the noise calculated in the KS system (dashed-dotted lines) deviates strongly from the reference noise, as it is expected for the unphysical KS system.
For the chosen parameters, the KS energy level is positioned at the Fermi energy, and the smeared-out step in the noise spectrum is thus removed in the KS system.
The dotted lines show noise spectra, which are derived from conductances, corrected with the HXC kernel of the \emph{adiabatic} HXC potential.
Here, the discrepancy with the perturbative result is reduced for very low frequencies.
From Eq.~\eqref{eq_fhxc} and Fig.~\ref{fig_fHXC} we estimate, that the frequency independent HXC kernel, $f_\mathrm{HXC}^{(\mathrm{A})}$, is reasonable for $\omega \ll 1/\tau_\mathrm{ch}$, i.\,e.~where
the frequency-dependent part of $f_\mathrm{HXC}^{(\mathrm{M})}$ is small.
On the contrary, the kernel of the \emph{non-adiabatic} HXC potential significantly improves the TDDFT result in the low and medium frequency regime (dashed lines).
The agreement with the reference noise is excellent for frequencies $\omega \lesssim T$, i.\,e.~for thermal noise, as it is visible in both plots in Fig.~\ref{fig_noise}.
We emphasize that already the linear frequency dependence of the associated HXC kernel, $f_\mathrm{HXC}^{(\mathrm{M})}$, is sufficient for the observed improvement.
The failure for frequencies $\omega > T$ has its root in the derivation of the related HXC potential, which was based on a Markov approximation \cite{Dittmann17}.

%%%%%%%%%%%%%%%%%%%%%%%%%%%%%%%%%%%%%%%%%%%%%%%%%%%%%%%%%%%%%%%%%%%%%%%%%%%%%%
\section{Conclusion}
\label{sec_conclusion}

We calculated noise spectra of an interacting mesoscopic capacitor in TDDFT using a recently developed non-adiabatic approximation for the HXC potential \cite{Dittmann17}. 
Thereby, we extended the discussion of our previous work to linear-response physics and found an HXC kernel with linear frequency dependence.  
The derived TDDFT noise agrees well with reference data from perturbation theory for thermal noise frequencies, $\omega \lesssim T$.
%%%%%%%%%%%%%%%%%%
\ack
Support from the Deutsche Forschungsgemeinschaft via RTG1995 (ND, JS) and an Emmy-Noether grant (NH), as well as 
from the Knut and Alice Wallenberg foundation and the Swedish VR (JS), is gratefully acknowledged.

%%%%%%%%%%%%%%%%%%%%%%%%%%%%%%%%%%%%%%%%%%%%%%%%%%%%%%%%%%%%%%%%%%%%%%%%%%%%%%%%%%%%%%%%%%%%%%%%%%%%%%%%%%%%%%%%%%%%%%%%%%%%%%%%%%%%%%%%%%%%%%%%%%%%%%%%%%%%
%\bibliographystyle{iopart-num}
%\bibliography{cite_c1,cite_c2,cite}

\providecommand{\newblock}{}

\end{document}